\documentclass[letters,fleqn, usenatbib]{mnras}

\usepackage{newtxtext,newtxmath}
\usepackage{txfonts}

\usepackage[T1]{fontenc}
\usepackage{ae,aecompl}

\usepackage{graphicx}	
\usepackage{amsmath}	
\usepackage{amssymb}	

\newcommand{\kms}{km\,s$^{-1}$}

\newcommand{\vsini}{$v\sin{i}$}

\title[Very weak magnetic field on Alhena]{Discovery of a very weak magnetic field on the Am star Alhena}

\author[Blaz\`ere et al.]{
A. Blaz\`ere,$^{1,2}$\thanks{E-mail: aurore.blazere@obspm.fr},
C. Neiner$^{1}$,
P. Petit$^{2,3}$
\\
$^{1}$LESIA, Observatoire de Paris, PSL Research University, CNRS, Sorbonne Universit\'es, UPMC Univ. Paris 06, Univ. Paris Diderot,\\ 
Sorbonne Paris Cit\'e, 5 place Jules Janssen, 92195 Meudon, France\\
$^{2}$Universit\'e de Toulouse, UPS-OMP, Institut de Recherche en Astrophysique et Plan\'etologie, Toulouse, France \\
$^{3}$CNRS, Institut de Recherche en Astrophysique et Plan\'etologie, 14 Avenue Edouard Belin, F-31400 Toulouse, France\\
}

\date{Accepted XXX. Received YYY; in original form ZZZ}

\pubyear{2016}

\begin{document}
\label{firstpage}
\pagerange{\pageref{firstpage}--\pageref{lastpage}}
\maketitle

\begin{abstract}
Alhena ($\gamma$ Gem) was observed in the frame of the BRITE (BRIght Target
Explorer) spectropolarimetric survey, which gathers high resolution, high
signal-to-noise, high sensitivity, spectropolarimetric observations of all stars
brighter than V=4 to combine seismic and spectropolarimetric studies of bright
stars.

We present here the discovery of a very weak magnetic field \textbf{on} the Am
star Alhena, thanks to very high signal-to-noise spectropolarimetric data
obtained with Narval at T\'elescope Bernard Lyot (TBL). All previously
studied Am stars show the presence of ultra-weak (sub-Gauss) fields with Zeeman
signatures with an unexpected prominent positive lobe. However, Alhena presents
a slightly stronger (but still very weak, only a few Gauss) field with normal
Zeeman signatures with a positive and negative lobe, as found in stronger field
(hundreds or thousands of Gauss)  stars. It is the first detection of a normal
magnetic signature in an Am star.

Alhena is thus a very interesting object, which might provide the clue to
understanding the peculiar shapes of the magnetic signatures of the other Am
stars.
\end{abstract}




\section{Introduction}

\subsection{Magnetism in hot stars}

Over the last decades, magnetic fields have been discovered in a significant
number of hot (A, B, and O) stars, and these fields probably play a significant
role in their evolution. However, the detailed properties of hot star magnetism
are not well understood yet. About 7\% of hot stars are found to be magnetic
\citep{grunhut15b} with dipolar magnetic fields above 300 G. The
detection rate for the A-type stars is similar
\citep[$\sim$10\%,][]{wolff68,power07}. In addition, sub-Gauss longitudinal
magnetic fields have recently been discovered in a few A and Am stars.

The normal A star Vega was the first ultra-weak field star discovered
\citep{lignieres09}. Its spectropolarimetric time series was interpreted in
terms of an ultra-weak surface magnetic field using Zeeman-Doppler Imaging
(ZDI).The results of this study support the fact that Vega is a rapidly
rotating star seen nearly pole-on. The reconstructed magnetic topology
revealed a magnetic region close to the pole with radial field orientation. 

In addition, ultra-weak magnetic field signatures have been detected in three Am
stars: Sirius\,A \citep{petit11}, $\beta$\,UMa, and $\theta$\,Leo
\citep{blazere16}, thanks to very precise spectropolarimetric observations.  For
these objects, the signature in circular polarization is not of null integral
over the line profile but exhibits a positive lobe
dominating over the negative one. This peculiar signal, although not expected in the standard Zeeman effect theory, was demonstrated to follow the same dependence on spectral line parameters as a signal of magnetic origin and  has been confirmed to be magnetic
\citep{blazere16}. Preliminary explanations are being proposed to explain the
peculiar shape of the signatures. In Am stars, high-resolution spectra have
revealed stronger microturbulence compared to normal A stars
\citep{landstreet09}. The very shallow convective shell producing this turbulent
velocity field may host supersonic convection flows \citep{kupka09}. This could
provide the source of sharp velocity and magnetic gradients needed to produce
strongly asymmetric profiles. Shocks traveling in this superficial turbulent
zone may also contribute to amplify any existing magnetic field.

Vega, Sirius\,A, $\beta$\,UMa, and $\theta$\,Leo may well be the first confirmed
members of a new class of magnetic hot stars: the ultra weakly magnetic hot
stars. Such ultra weak magnetic fields are difficult to detect due to
the weak amplitude of their Zeeman signatures and may exist in other
hot stars. However, ultra weakly magnetic stars are considered a
separate class of magnetic stars compared to the $\sim$7\% of stronger field
stars, because no magnetic stars exist with a polar field strength between
$\sim$300 G and the Gauss-level fields observed in ultra weak magnetic stars.

To explain this dichotomy between strong and weak magnetic fields in
hot stars, \cite{auriere07} proposed a new scenario based on the stability of a
large scale magnetic configuration in a differentially rotating star: strong
magnetic fields correspond to stable configurations and weak magnetic fields to
unstable configurations. Another theory to explain the dichotomy is the failed
fossil theory \citep{braithwaite13}: strong magnetic fields rapidly reach an
equilibrium whereas weak magnetic fields are still dynamically evolving towards
the equilibrium and decreased due to the instability.

\subsection{The Am star Alhena}

Alhena was observed in the frame of the BRITE (BRIght Target Explorer)
spectropolarimetric survey. The BRITE constellation of nano-satellites performs
asteroseismology of stars with V$\leq$4 \citep{weiss14}. In this context, we
are performing a high resolution, high signal-to-noise (S/N), high sensitivity,
spectropolarimetric survey of all stars brighter than V=4, with the ultimate aim
to combine seismic and spectropolarimetric studies of bright stars.

Alhena is a bright (V=1.90) spectroscopic binary, in which the primary is a
subgiant A0IVm star \citep{gray14} and the secondary is a cool G star
\citep{thalmann14}. The orbital elements of the binary have been measured thanks
to interferometry \citep{drummond14}. The orbital period is 12.63 years and the
orbit is very eccentric with $e$=0.89. The mass of the primary is estimated to
2.84 M$_{\odot}$ and the one of the secondary to 1.07 M${\odot}$. The primary,
Alhena\,A, is a weakly Am star \citep{adelman15}, similar to the three known 
magnetic Am stars. The stellar parameters of Alhena\,A are actually very close
to the ones of $\theta$\,Leo, which exhibits peculiar magnetic signatures in its
Stokes V profiles (see Table~\ref{param}).

\begin{table}
\caption{Fundamental parameters of the Am stars Alhena and $\theta$ Leo.}
\begin{tabular}{l l l}
\hline
  & Alhena\,A & $\theta$\,Leo\\
\hline
\hline
   Spectral type & A0IVm & A2Vm \\
   $T_{\rm{eff}}$   (K)  & 9260$\pm$10$^{a}$ & 9280$\pm$10$^{b}$  \\
   log g   &   3.6$^a$   & 3.65$^c$ \\
   Mass ($M_{\odot}$) & 2.84$\pm$0.01$^{b}$ & 2.94$\pm$0.2$^{b}$\\
   Radius ($R_{\odot}$) & 3.9$\pm$0.1 $^{d}$ & 4.03$\pm$ 0.10 $^{e}$\\
   \vsini \ (\kms) & 15$\pm$3$^{f}$ & 23$\pm$3$^{f}$\\
   Luminosity (L$_{\odot}$) & 123$\pm$11$^{b}$ & 127$\pm$13$^{b}$\\
   Age (Myr) & 484$^{g}$ & 436$^{g}$ \\
   Microturb. (\kms) & 2$^{a}$ & 1$^{b}$\\
\hline

$^{a}$ \cite{adelman15} & \multicolumn{2}{l}{$^{b}$ \cite{zorec12}}\\
$^{c}$ \cite{adelman15b} & \multicolumn{2}{l}{$^{d}$ \cite{pasinetti01}}\\
$^{e}$ \cite{royer07} & \multicolumn{2}{l}{$^{f}$ \cite{royer02}}\\
$^{h}$ \cite{david15}\\

\end{tabular}
\label{param}
\end{table}

\section{Observations}

Data were collected with the NARVAL spectropolarimeter (\citealt{auriere03},
\citealt{silvester12}), installed at the 2-meter Bernard Lyot Telescope (TBL)
at the summit of Pic du Midi Observatory in the French Pyr\'en\'ees. Narval is a
high-resolution spectropolarimeter, very efficient to
detect stellar magnetic fields thanks the polarization they generate in
photospheric spectral lines. It covers a wavelength domain from about 375 to
1050 nm, with a resolving power of $\sim$ 68000.

We used the polarimetry mode to measure the circular polarization (Stokes V).
The 4 sub-exposures are constructively combined to obtain the Stokes V spectrum
in addition to the intensity (Stokes I) spectrum. The sub-exposures are also
destructively combined to produce a null polarization (N) spectrum to check for
spurious detection due to e.g., instrumental effects, variable observing
conditions, or non-magnetic physical effects such as pulsations. Alhena was
observed on October 27, 2014, and September 18 and 19, 2015. The journal of
observations is provided in Table~\ref{journal}.

\begin{table}
\label{journal}
\caption{Journal of observations indicating the Heliocentric Julian Date at the
middle of the observations (mid-HJD - 2450000), the exposure time in seconds,
and the mean signal-to-noise ratio S/N of the intensity spectrum at $\sim$ 500 nm.}
\centering
\begin{tabular}{l l l l}
\hline
  & 27 Oct 2014 & 18 Sep 2015 & 19 Sep 2015\\
\hline
\hline
mid-HJD & 6958.650 & 7284.695 & 7285.694\\
$T_{\rm exp}$ (s) & 4$\times$25 & 4$\times$35 & 4$\times$35 \\
S/N & 986 & 1016 & 1093\\
\hline
\end{tabular}
\end{table} 
 
We used the Libre-Esprit reduction package \citep{donati97} to reduce the data.
We then normalized each of the 40 echelle orders of each of the 3 Stokes I
spectra with the continuum task of IRAF\footnote{IRAF is distributed by the
National Optical Astronomy Observatories, which are operated by the Association
of Universities for Research in Astronomy, Inc., under cooperative agreement
with the National Science Foundation.}. We applied the same normalization to the
Stokes V and N spectra.

\section{Magnetic analysis}

To test whether Alhena is magnetic, we use the Least Square Deconvolution (LSD)
technique. It is a cross-correlation technique for computing average pseudo-line
profiles from a list of spectral lines in order to increase the S/N ratio. Under
several rough approximations (additive line profiles, wavelength independent
limb-darkening, self-similar local profile shape, weak magnetic fields), stellar
spectra can indeed be seen as a line pattern convolved with an average line
profile.

We first created a line mask corresponding to the primary component of Alhena.
We started from a list of lines extracted from VALD \citep{piskunov95, kupka99}
for an A star with $T_{\rm eff}$=9250 K and log g=3.5, with their Land\'e
factors and theoretical line depths. We then cleaned this line list by removing
the hydrogen lines, the lines that are blended with hydrogen lines, as well as
those that are not visible in the spectra. We also added some lines visible in
the spectra that were not in the original A-star mask. Altogether we obtained a
mask of 1052 lines. We then adjusted the depth of these 1052 lines in the mask
to fit the observed line depths.

\begin{figure}
\resizebox{\hsize}{!}{\includegraphics[trim=0.5cm 0.5cm 4.5cm 3.2cm, clip]{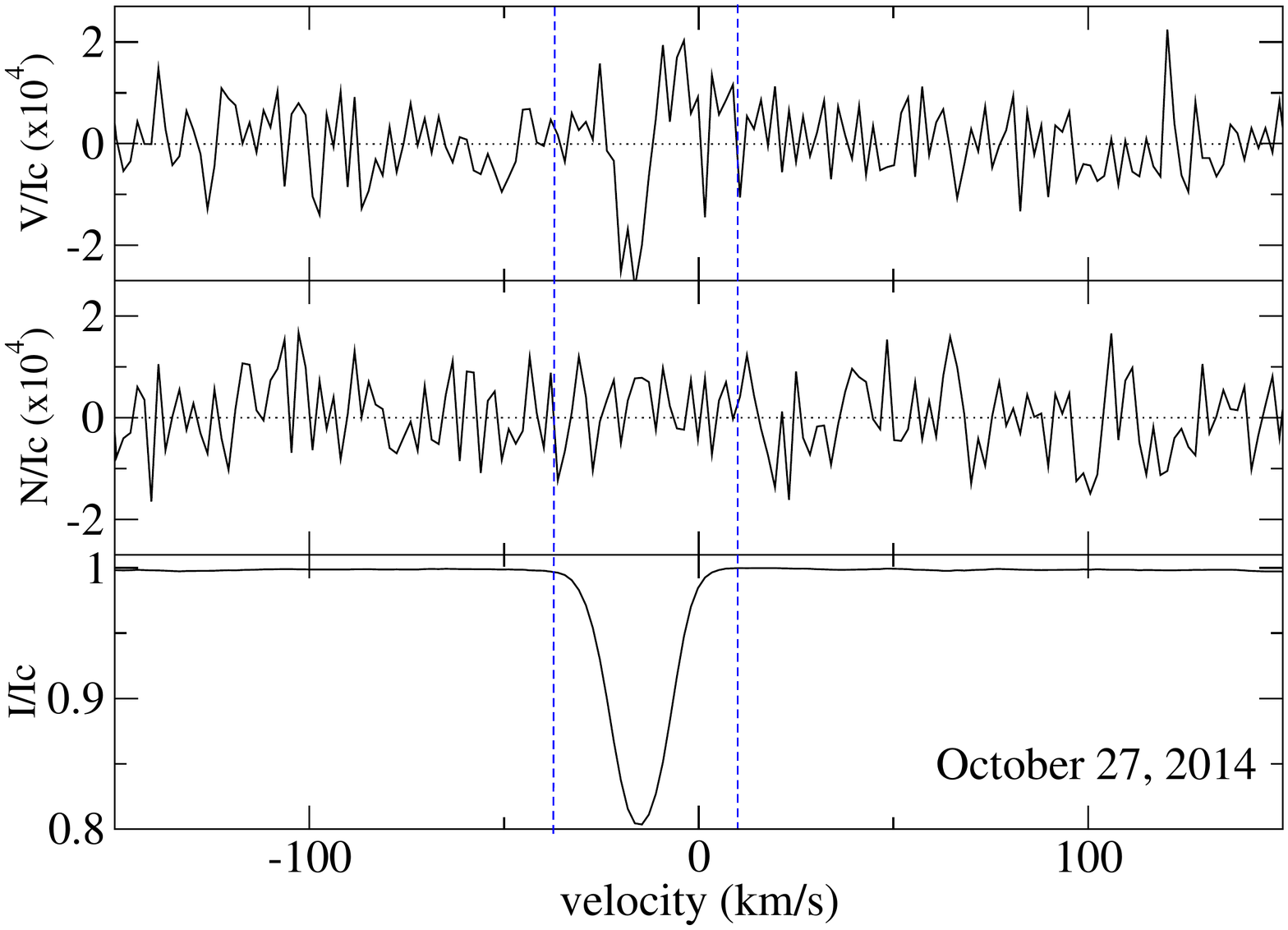}}
\resizebox{\hsize}{!}{\includegraphics[trim=0.5cm 0.5cm 4.5cm 3.2cm, clip]{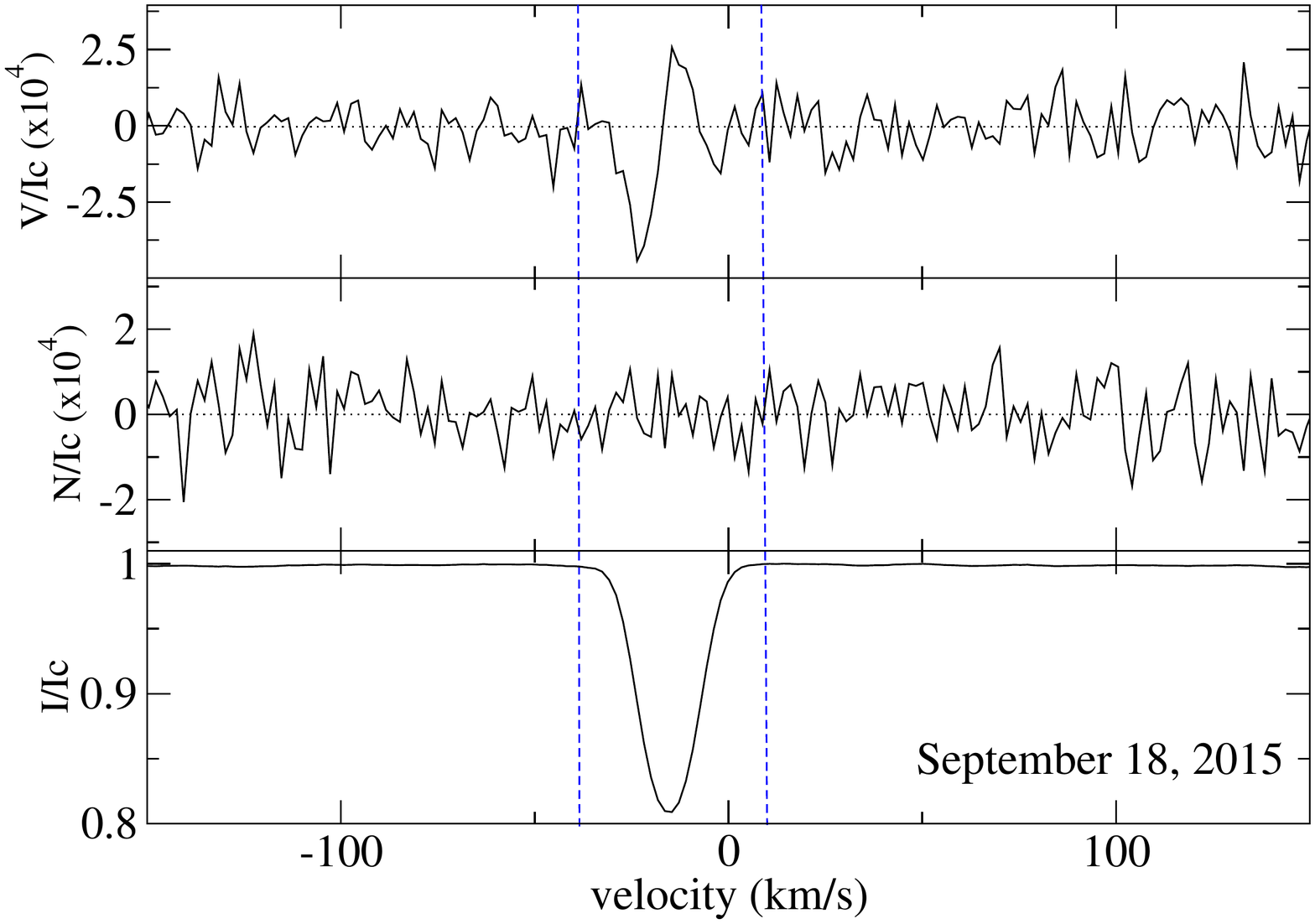}}
\resizebox{\hsize}{!}{\includegraphics[trim=0.5cm 0.5cm 4.5cm 3.2cm, clip]{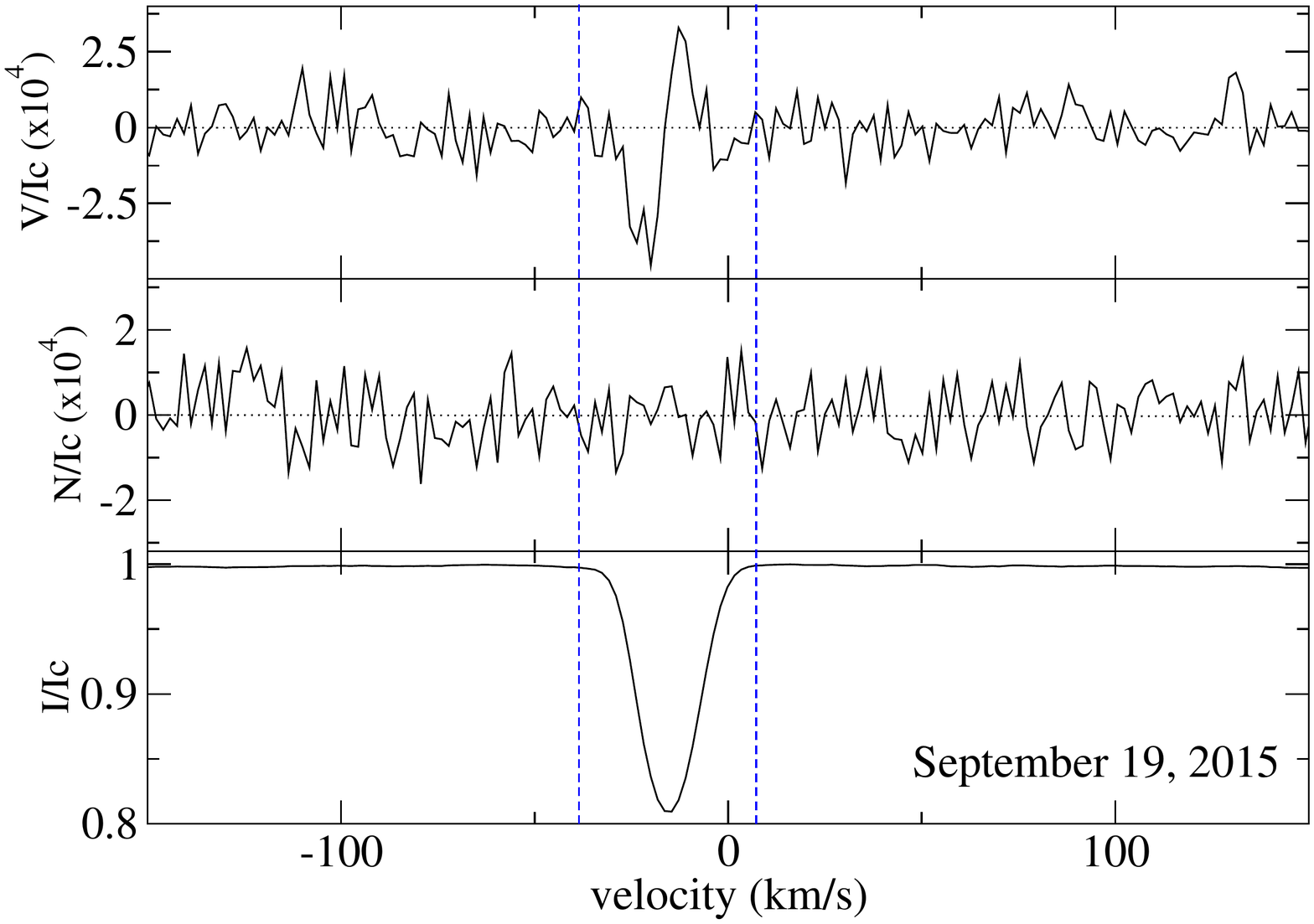}}
\caption{LSD profiles of Alhena\,A for Stokes I, the null N
polarization, and
Stokes V for the October 27, 2014 (top), September 18, 2015 (center) and
September 19, 2015 (bottom). The vertical blue dashed lines indicate the
domain of integration used to determine the longitudinal field values.}
\label{lsd}
\end{figure}

The results of the LSD analysis are show in  Fig.~\ref{lsd}. The Stokes V
profiles show clear Zeeman signatures for all 3 nights. We computed the
detection probability of the Stokes V signal by using the $\chi^2$ test proposed
by \cite{donati92}, getting a detection probability of $\approx$100\% for all
observations, with a false alarm probability always smaller than
1.117$\times$10$^{-7}$ inside the stellar line. Therefore, we have a definite
detection of a magnetic field in all 3 nights. We note that the magnetic
signatures are strong enough to be detected in individual LSD Stokes V profiles,
while for other magnetic Am stars co-addition of many Stokes V profiles were
necessary to extract a magnetic signature \citep{blazere16}. Outside the stellar
line, we obtained a detection probability between 20\% and 40\% and a false
alarm probability between 7.356$\times$10$^{-1}$ and 5.351$\times$10$^{1}$, that
corresponds to a non detection outside the stellar lines. 

\begin{figure}
\resizebox{\hsize}{!}{\includegraphics[trim=0.5cm 1cm 4.5cm 3cm, clip]{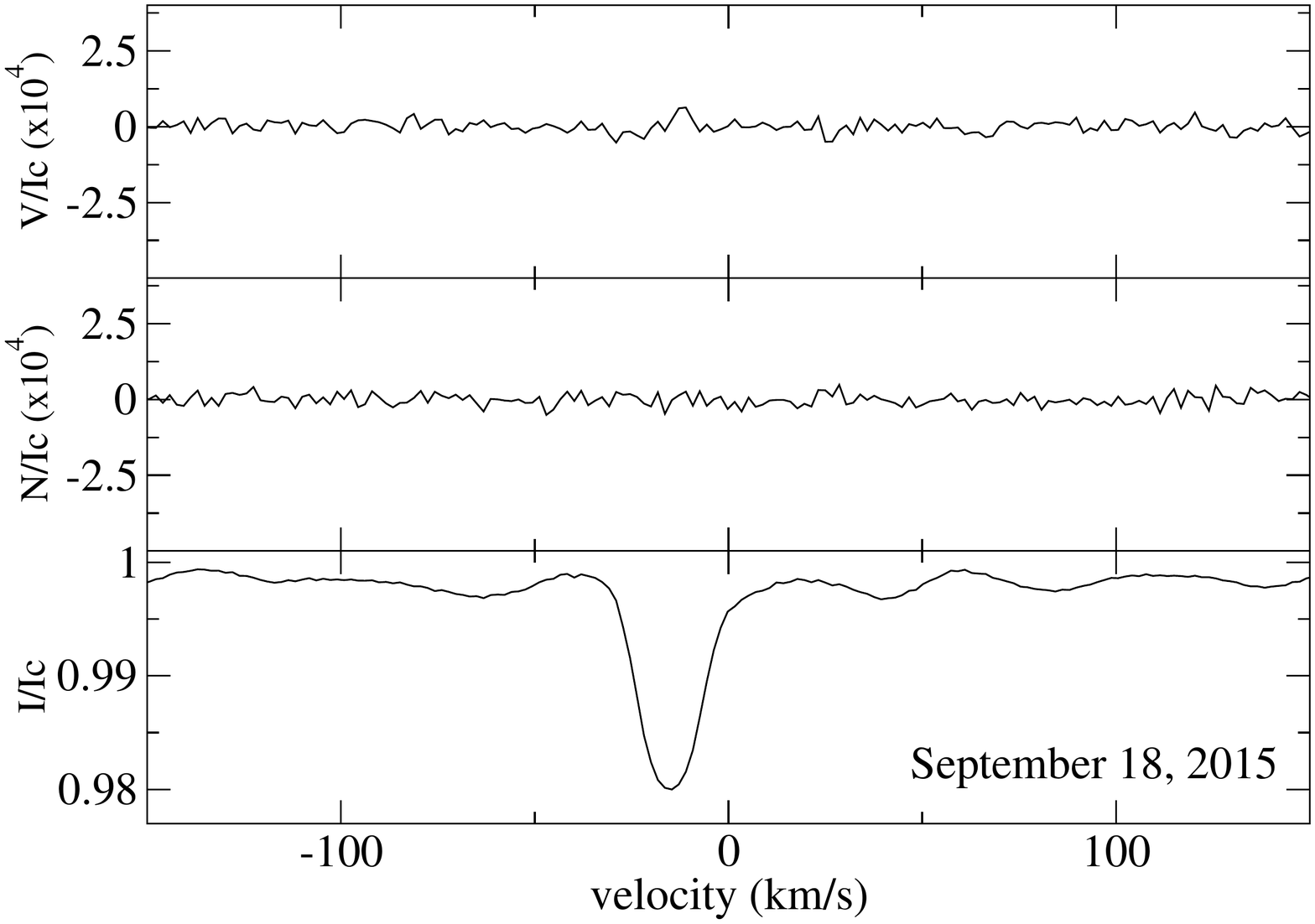}}
\caption{LSD profiles of the companion Alhena\,B for Stokes I, the null N
polarization, and Stokes V for September 18, 2015.}
\label{lsdsec}
\end{figure}

Since the diameter of the fibre of Narval is 2.8 arcsec, the two components of
the binary have been recorded in the observations. However, the secondary is 5-6
magnitudes fainter than the primary, so only $\sim$2\% of the received light
comes from the secondary component. Moreover, the secondary is not visible in
the spectra. Thus, the contribution of the lines of the secondary are considered negligible unless its radial velocity is very close to the one of the primary.
In addition, we ran the LSD analysis with a mask corresponding to a main
sequence G star and the signatures in the Stokes V profiles disappeared (see
Fig.~\ref{lsdsec}). We thus confirm that the signatures in the Stokes V profiles
come from the primary star, i.e. that the Am star is magnetic.

Using the centre-of-gravity method \citep{rees79} with a mean wavelength of 500
nm and a mean Land\'e factor of $\sim$ 1.46 corresponding to the normalization
parameters used in the LSD, we calculated the longitudinal field value ($B_l$)
corresponding to these Zeeman signatures over the velocity range [-40:8] \kms. 

\begin{equation}
B_{l} \propto -\frac{\int v V(v)dv}{\lambda_{0}g_{m}c \int (1-I(v))dv}
\end{equation}

where v (\kms) is the radial velocity, $\lambda_{0}$ (nm) the normalized
wavelength of the line-list used to compute the LSD profiles, g the normalized
Land\'e factor and c (\kms) the light velocity. The longitudinal magnetic field
value for the three observations, and the corresponding null values, are shown
in Table~\ref{bl}. The values of the longitudinal magnetic field is around $-5$
G, with an error bar smaller than 3 G. The values extracted from the N profiles
are compatible with 0 G.

\begin{table}
\label{bl}
\caption{Longitudinal magnetic field (B$_l$) and null (N) measurements with their respective error bars and magnetic detection status.}
\centering
\begin{tabular}{c c c c}
\hline
  & Oct. 27, 2014 & Sep. 18, 2015 & Sep. 19, 2015\\
\hline
\hline
B$_l$ (G)& -5.1$\pm$2.7 & -5.6$\pm$2.7 & -5.5$\pm$2.5 \\
N (G) & -1.5$\pm$2.7 & 1.6$\pm$2.7 &-2.6$\pm$2.5\\
Detection & Definite & Definite & Definite\\
\hline
\end{tabular}
 \end{table}

The shape of the Zeeman signatures in the Stokes V profiles slightly changed
between the observation obtained in 2014 and the ones from 2015. This could be
due to a rotational modulation of the longitudinal magnetic field, if the field
is oblique compared to the rotation axis, as observed in most hot stars
\citep{grunhut15b}. In 2014, the signature  would look like a
cross-over signature, while in 2015 the negative pole may be observed.

On the contrary, the signatures obtained over two consecutive nights in 2015 did
not change. This suggests that either Alhena is an intrinsically slow rotator,
or the rotational modulation is small because the star is seen under a specific
geometrical configuration with an inclination or obliquity angle close to 0. 

\section{Discussion and conclusion}

The observations presented in this paper correspond to the first detection of a
magnetic field in an Am star with a normal Zeeman signature, i.e. with a
positive and negative lobe as seen in the ultra-weakly magnetic A star Vega and
in all strongly magnetic hot stars. On the contrary, all the other Am stars
studied in spectropolarimetry with a high accuracy exhibit peculiar magnetic
signatures with only a prominent positive lobe.

The difference between the field of Alhena and the other Am stars is thus
puzzling. In particular, Alhena has very similar stellar parameters as the ones
of the magnetic Am star $\theta$\,Leo. However, the signatures in the Stokes V
profiles are very different. $\theta$\,Leo shows peculiar signatures, while
Alhena shows normal signatures. Considering the oblique rotator model,
the dipolar magnetic field $B_{d}$ is at least 3.3 times the maximum observed
$B_{l}$ value \citep{preston67,auriere07}. Therefore, the longitudinal field
values measured for Alhena point towards a polar magnetic field strength of the
order of 15 G, i.e. weak but much stronger than what is observed in
$\theta$\,Leo and the other magnetic Am stars.

An explanation to the difference between the characteristics of the magnetic
field observed in Alhena and in other Am stars may be found in their
microturbulence value. The microturbulence of Alhena\,A is $\sim1$ \kms
\citep{adelman15}, while the one of $\theta$\,Leo, $\beta$\,UMa, and Sirius\,A
is $\sim2$ \kms \citep{adelman15b,adelman11,landstreet09}. Indeed, the peculiar
shape of the magnetic signatures of the latter 3 Am stars is thought to be
related to their stronger microturbulence, compared to normal A stars
\citep{blazere16}. The very shallow convective shell producing this turbulent
velocity field may host supersonic convection flows \citep{kupka09}, which could
be the source of sharp velocity and magnetic gradients producing strongly
asymmetric Zeeman profiles. Alhena\,A may have a too weak microturbulence to
undergo this effect.

Another difference between Alhena and $\theta$\,Leo is that Alhena is a binary
star with a G-type companion, while $\theta$\,Leo is a single star. However,
Sirius is also a binary star and Sirius\,A does present peculiar magnetic
signatures like $\theta$\,Leo. The distance between the two components
of Sirius is larger ($P_{\rm orb}$=50.1 years) than the one of Alhena,
nevertheless Alhena is a wide binary as well ($P_{\rm orb}$=12.63 years).
However, the orbit of Alhena B is more eccentric ($e$=0.89) than the orbit of
Sirius B ($e$=0.59), and there could thus be more tidal interactions between the
2 components of Alhena than between the components of Sirius.

Alhena is thus a very interesting star to understand the magnetism of Am stars
and ultra-weak magnetic fields in general. We will continue to observe Alhena in
the frame of the BRITE spectropolarimetric survey to obtain more information
about its magnetic field. In particular, the comparison between the
observations obtained in 2014 and 2015 indicate that the Stokes V profile could
be rotationally modulated, although either the rotation period is long or the
geometrical configuration leads to only weak modulation. This could be tested,
and the geometrical configuration constrained, thanks to more
spectropolarimetric observations spread over the rotation period of Alhena\,A. 

\section*{Acknowledgements}
We thank Oleg Kochukhov and Colin Folsom for useful discussions.
We acknowledge support from the ANR (Agence Nationale de la Recherche) project
Imagine. This research has made use of the SIMBAD database operated at CDS,
Strasbourg (France), and of NASA\'\ s Astrophysics Data System (ADS).




\bibliographystyle{mnras}
\bibliography{biblio_v2} 

\begin{thebibliography}{}
\makeatletter
\relax
\def\mn@urlcharsother{\let\do\@makeother \do\$\do\&\do\#\do\^\do\_\do\%\do\~}
\def\mn@doi{\begingroup\mn@urlcharsother \@ifnextchar [ {\mn@doi@}
  {\mn@doi@[]}}
\def\mn@doi@[#1]#2{\def\@tempa{#1}\ifx\@tempa\@empty \href
  {http://dx.doi.org/#2} {doi:#2}\else \href {http://dx.doi.org/#2} {#1}\fi
  \endgroup}
\def\mn@eprint#1#2{\mn@eprint@#1:#2::\@nil}
\def\mn@eprint@arXiv#1{\href {http://arxiv.org/abs/#1} {{\tt arXiv:#1}}}
\def\mn@eprint@dblp#1{\href {http://dblp.uni-trier.de/rec/bibtex/#1.xml}
  {dblp:#1}}
\def\mn@eprint@#1:#2:#3:#4\@nil{\def\@tempa {#1}\def\@tempb {#2}\def\@tempc
  {#3}\ifx \@tempc \@empty \let \@tempc \@tempb \let \@tempb \@tempa \fi \ifx
  \@tempb \@empty \def\@tempb {arXiv}\fi \@ifundefined
  {mn@eprint@\@tempb}{\@tempb:\@tempc}{\expandafter \expandafter \csname
  mn@eprint@\@tempb\endcsname \expandafter{\@tempc}}}

\bibitem[\protect\citeauthoryear{{Adelman}, {Yu}  \& {Gulliver}}{{Adelman}
  et~al.}{2011}]{adelman11}
{Adelman} S.~J.,  {Yu} K.,   {Gulliver} A.~F.,  2011, \mn@doi [Astronomische
  Nachrichten] {10.1002/asna.201011488}, \href
  {http://adsabs.harvard.edu/abs/2011AN....332..153A} {332, 153}

\bibitem[\protect\citeauthoryear{{Adelman}, {Gulliver}  \& {Heaton}}{{Adelman}
  et~al.}{2015a}]{adelman15b}
{Adelman} S.~J.,  {Gulliver} A.~F.,   {Heaton} R.~J.,  2015a, \mn@doi [\pasp]
  {10.1086/679636}, \href {http://adsabs.harvard.edu/abs/2015PASP..127...58A}
  {127, 58}

\bibitem[\protect\citeauthoryear{{Adelman}, {Gulliver}  \&
  {Kaewkornmaung}}{{Adelman} et~al.}{2015b}]{adelman15}
{Adelman} S.~J.,  {Gulliver} A.~F.,   {Kaewkornmaung} P.,  2015b, \mn@doi
  [\pasp] {10.1086/680998}, \href
  {http://cdsads.u-strasbg.fr/abs/2015PASP..127..340A} {127, 340}

\bibitem[\protect\citeauthoryear{{Auri{\`e}re}}{{Auri{\`e}re}}{2003}]{auriere03}
{Auri{\`e}re} M.,  2003, in {Arnaud} J.,  {Meunier} N.,  eds,  EAS Publications
  Series Vol. 9, EAS Publications Series. p.~105

\bibitem[\protect\citeauthoryear{{Auri{\`e}re} et~al.,}{{Auri{\`e}re}
  et~al.}{2007}]{auriere07}
{Auri{\`e}re} M.,  et~al., 2007, \mn@doi [\aap] {10.1051/0004-6361:20078189},
  \href {http://adsabs.harvard.edu/abs/2007A%26A...475.1053A} {475, 1053}

\bibitem[\protect\citeauthoryear{{Blaz{\`e}re} et~al.,}{{Blaz{\`e}re}
  et~al.}{2016}]{blazere16}
{Blaz{\`e}re} A.,  et~al., 2016, preprint, \href
  {http://adsabs.harvard.edu/abs/2016arXiv160101829B} {} (\mn@eprint {arXiv}
  {1601.01829})

\bibitem[\protect\citeauthoryear{{Braithwaite} \& {Cantiello}}{{Braithwaite} \&
  {Cantiello}}{2013}]{braithwaite13}
{Braithwaite} J.,  {Cantiello} M.,  2013, \mn@doi [\mnras]
  {10.1093/mnras/sts109}, \href
  {http://adsabs.harvard.edu/abs/2013MNRAS.428.2789B} {428, 2789}

\bibitem[\protect\citeauthoryear{{David} \& {Hillenbrand}}{{David} \&
  {Hillenbrand}}{2015}]{david15}
{David} T.~J.,  {Hillenbrand} L.~A.,  2015, \mn@doi [\apj]
  {10.1088/0004-637X/804/2/146}, \href
  {http://cdsads.u-strasbg.fr/abs/2015ApJ...804..146D} {804, 146}

\bibitem[\protect\citeauthoryear{{Donati}, {Semel}  \& {Rees}}{{Donati}
  et~al.}{1992}]{donati92}
{Donati} J.-F.,  {Semel} M.,   {Rees} D.~E.,  1992, \aap, \href
  {http://adsabs.harvard.edu/abs/1992A%26A...265..669D} {265, 669}

\bibitem[\protect\citeauthoryear{{Donati}, {Semel}, {Carter}, {Rees}  \&
  {Collier Cameron}}{{Donati} et~al.}{1997}]{donati97}
{Donati} J.-F.,  {Semel} M.,  {Carter} B.~D.,  {Rees} D.~E.,   {Collier
  Cameron} A.,  1997, \mnras, \href
  {http://adsabs.harvard.edu/abs/1997MNRAS.291..658D} {291, 658}

\bibitem[\protect\citeauthoryear{{Drummond}}{{Drummond}}{2014}]{drummond14}
{Drummond} J.~D.,  2014, \mn@doi [\aj] {10.1088/0004-6256/147/3/65}, \href
  {http://cdsads.u-strasbg.fr/abs/2014AJ....147...65D} {147, 65}

\bibitem[\protect\citeauthoryear{{Gray}}{{Gray}}{2014}]{gray14}
{Gray} D.~F.,  2014, \mn@doi [\aj] {10.1088/0004-6256/147/4/81}, \href
  {http://cdsads.u-strasbg.fr/abs/2014AJ....147...81G} {147, 81}

\bibitem[\protect\citeauthoryear{{Grunhut} \& {Neiner}}{{Grunhut} \&
  {Neiner}}{2015}]{grunhut15b}
{Grunhut} J.~H.,  {Neiner} C.,  2015, in {Nagendra} K.~N.,  {Bagnulo} S.,
  {Centeno} R.,   {Jes{\'u}s Mart{\'{\i}}nez Gonz{\'a}lez} M.,  eds,  IAU
  Symposium Vol. 305, IAU Symposium. p.~53

\bibitem[\protect\citeauthoryear{{Kupka} \& {Ryabchikova}}{{Kupka} \&
  {Ryabchikova}}{1999}]{kupka99}
{Kupka} F.,  {Ryabchikova} T.~A.,  1999, Publications de l'Observatoire
  Astronomique de Beograd, \href
  {http://adsabs.harvard.edu/abs/1999POBeo..65..223K} {65, 223}

\bibitem[\protect\citeauthoryear{{Kupka}, {Ballot}  \& {Muthsam}}{{Kupka}
  et~al.}{2009}]{kupka09}
{Kupka} F.,  {Ballot} J.,   {Muthsam} H.~J.,  2009, Communications in
  Asteroseismology, \href {http://adsabs.harvard.edu/abs/2009CoAst.160...30K}
  {160, 30}

\bibitem[\protect\citeauthoryear{{Landstreet}, {Kupka}, {Ford}, {Officer},
  {Sigut}, {Silaj}, {Strasser}  \& {Townshend}}{{Landstreet}
  et~al.}{2009}]{landstreet09}
{Landstreet} J.~D.,  {Kupka} F.,  {Ford} H.~A.,  {Officer} T.,  {Sigut}
  T.~A.~A.,  {Silaj} J.,  {Strasser} S.,   {Townshend} A.,  2009, \mn@doi
  [\aap] {10.1051/0004-6361/200912083}, \href
  {http://adsabs.harvard.edu/abs/2009A%26A...503..973L} {503, 973}

\bibitem[\protect\citeauthoryear{{Ligni{\`e}res}, {Petit}, {B{\"o}hm}  \&
  {Auri{\`e}re}}{{Ligni{\`e}res} et~al.}{2009}]{lignieres09}
{Ligni{\`e}res} F.,  {Petit} P.,  {B{\"o}hm} T.,   {Auri{\`e}re} M.,  2009,
  \mn@doi [\aap] {10.1051/0004-6361/200911996}, \href
  {http://adsabs.harvard.edu/abs/2009A%26A...500L..41L} {500, L41}

\bibitem[\protect\citeauthoryear{{Pasinetti Fracassini}, {Pastori}, {Covino}
  \& {Pozzi}}{{Pasinetti Fracassini} et~al.}{2001}]{pasinetti01}
{Pasinetti Fracassini} L.~E.,  {Pastori} L.,  {Covino} S.,   {Pozzi} A.,  2001,
  \mn@doi [\aap] {10.1051/0004-6361:20000451}, \href
  {http://cdsads.u-strasbg.fr/abs/2001A%26A...367..521P} {367, 521}

\bibitem[\protect\citeauthoryear{{Petit} et~al.,}{{Petit}
  et~al.}{2011}]{petit11}
{Petit} P.,  et~al., 2011, \mn@doi [\aap] {10.1051/0004-6361/201117573}, \href
  {http://adsabs.harvard.edu/abs/2011A%26A...532L..13P} {532, L13}

\bibitem[\protect\citeauthoryear{{Piskunov}, {Kupka}, {Ryabchikova}, {Weiss}
  \& {Jeffery}}{{Piskunov} et~al.}{1995}]{piskunov95}
{Piskunov} N.~E.,  {Kupka} F.,  {Ryabchikova} T.~A.,  {Weiss} W.~W.,
  {Jeffery} C.~S.,  1995, \aaps, \href
  {http://adsabs.harvard.edu/abs/1995A%26AS..112..525P} {112, 525}

\bibitem[\protect\citeauthoryear{{Power}, {Wade}, {Hanes}, {Aurier}  \&
  {Silvester}}{{Power} et~al.}{2007}]{power07}
{Power} J.,  {Wade} G.~A.,  {Hanes} D.~A.,  {Aurier} M.,   {Silvester} J.,
  2007, in {Romanyuk} I.~I.,  {Kudryavtsev} D.~O.,  {Neizvestnaya} O.~M.,
  {Shapoval} V.~M.,  eds, Physics of Magnetic Stars. pp 89--97 (\mn@eprint {}
  {astro-ph/0612557})

\bibitem[\protect\citeauthoryear{{Rees} \& {Semel}}{{Rees} \&
  {Semel}}{1979}]{rees79}
{Rees} D.~E.,  {Semel} M.~D.,  1979, \aap, \href
  {http://adsabs.harvard.edu/abs/1979A%26A....74....1R} {74, 1}

\bibitem[\protect\citeauthoryear{{Royer}, {Gerbaldi}, {Faraggiana}  \&
  {G{\'o}mez}}{{Royer} et~al.}{2002}]{royer02}
{Royer} F.,  {Gerbaldi} M.,  {Faraggiana} R.,   {G{\'o}mez} A.~E.,  2002,
  \mn@doi [\aap] {10.1051/0004-6361:20011422}, \href
  {http://adsabs.harvard.edu/abs/2002A%26A...381..105R} {381, 105}

\bibitem[\protect\citeauthoryear{{Royer}, {Zorec}  \& {G{\'o}mez}}{{Royer}
  et~al.}{2007}]{royer07}
{Royer} F.,  {Zorec} J.,   {G{\'o}mez} A.~E.,  2007, \mn@doi [\aap]
  {10.1051/0004-6361:20065224}, \href
  {http://cdsads.u-strasbg.fr/abs/2007A%26A...463..671R} {463, 671}

\bibitem[\protect\citeauthoryear{{Silvester}, {Wade}, {Kochukhov}, {Bagnulo},
  {Folsom}  \& {Hanes}}{{Silvester} et~al.}{2012}]{silvester12}
{Silvester} J.,  {Wade} G.~A.,  {Kochukhov} O.,  {Bagnulo} S.,  {Folsom} C.~P.,
    {Hanes} D.,  2012, \mn@doi [\mnras] {10.1111/j.1365-2966.2012.21587.x},
  \href {http://adsabs.harvard.edu/abs/2012MNRAS.426.1003S} {426, 1003}

\bibitem[\protect\citeauthoryear{{Thalmann} et~al.,}{{Thalmann}
  et~al.}{2014}]{thalmann14}
{Thalmann} C.,  et~al., 2014, \mn@doi [\aap] {10.1051/0004-6361/201424581},
  \href {http://cdsads.u-strasbg.fr/abs/2014A%26A...572A..91T} {572, A91}

\bibitem[\protect\citeauthoryear{{Weiss} et~al.,}{{Weiss}
  et~al.}{2014}]{weiss14}
{Weiss} W.~W.,  et~al., 2014, in {Guzik} J.~A.,  {Chaplin} W.~J.,  {Handler}
  G.,   {Pigulski} A.,  eds,  IAU Symposium Vol. 301, IAU Symposium. p.~67

\bibitem[\protect\citeauthoryear{{Wolff}}{{Wolff}}{1968}]{wolff68}
{Wolff} S.~C.,  1968, \mn@doi [\pasp] {10.1086/128630}, \href
  {http://adsabs.harvard.edu/abs/1968PASP...80..281W} {80, 281}

\bibitem[\protect\citeauthoryear{{Zorec} \& {Royer}}{{Zorec} \&
  {Royer}}{2012}]{zorec12}
{Zorec} J.,  {Royer} F.,  2012, \mn@doi [\aap] {10.1051/0004-6361/201117691},
  \href {http://adsabs.harvard.edu/abs/2012A%26A...537A.120Z} {537, A120}

\makeatother
\end{thebibliography}


\bsp	
\label{lastpage}
\end{document}